\documentclass[reprint,aps,prc,superscriptaddress]{revtex4-2}

\usepackage[margin=1in]{geometry}
\usepackage{graphicx}
\usepackage{hyperref}
\usepackage{slashed}
\usepackage[caption=false]{subfig}

\def\xf{$^{134}$Xe}
\def\xs{$^{136}$Xe}
\def\tntb{$2\nu2\beta$}
\def\zntb{$0\nu2\beta$}
\def\ttn{$T_{1/2}^{2\nu}$}
\def\tzn{$T_{1/2}^{0\nu}$}
\def\rz{$^{220}$Rn}
\def\rt{$^{222}$Rn}
\def\ke{$^{85}$Kr}

\begin{document}
\title{Projected sensitivity of the LUX-ZEPLIN (LZ) experiment to the two-neutrino and neutrinoless double beta decays of \xf}
\author{D.S.~Akerib}
\affiliation{SLAC National Accelerator Laboratory, Menlo Park, CA 94025-7015, USA}
\affiliation{Kavli Institute for Particle Astrophysics and Cosmology, Stanford University, Stanford, CA  94305-4085 USA}

\author{A.K.~Al~Musalhi}
\affiliation{University of Oxford, Department of Physics, Oxford OX1 3RH, UK}

\author{S.K.~Alsum}
\affiliation{University of Wisconsin-Madison, Department of Physics, Madison, WI 53706-1390, USA}

\author{C.S.~Amarasinghe}
\affiliation{University of Michigan, Randall Laboratory of Physics, Ann Arbor, MI 48109-1040, USA}

\author{A.~Ames}
\affiliation{SLAC National Accelerator Laboratory, Menlo Park, CA 94025-7015, USA}
\affiliation{Kavli Institute for Particle Astrophysics and Cosmology, Stanford University, Stanford, CA  94305-4085 USA}

\author{T.J.~Anderson}
\affiliation{SLAC National Accelerator Laboratory, Menlo Park, CA 94025-7015, USA}
\affiliation{Kavli Institute for Particle Astrophysics and Cosmology, Stanford University, Stanford, CA  94305-4085 USA}

\author{N.~Angelides}
\affiliation{University College London (UCL), Department of Physics and Astronomy, London WC1E 6BT, UK}

\author{H.M.~Ara\'{u}jo}
\affiliation{Imperial College London, Physics Department, Blackett Laboratory, London SW7 2AZ, UK}

\author{J.E.~Armstrong}
\affiliation{University of Maryland, Department of Physics, College Park, MD 20742-4111, USA}

\author{M.~Arthurs}
\affiliation{University of Michigan, Randall Laboratory of Physics, Ann Arbor, MI 48109-1040, USA}

% ...

\collaboration{The LUX-ZEPLIN (LZ) Collaboration}

\date{\today}
\begin{abstract}
  \noindent The projected sensitivity of the LUX-ZEPLIN (LZ) experiment to two-neutrino and neutrinoless double beta decay of \xf\ is presented. LZ is a 10-tonne xenon time projection chamber optimized for the detection of dark matter particles, that is expected to start operating in 2021 at Sanford Underground Research Facility, USA. Its large mass of natural xenon provides an exceptional opportunity to search for the double beta decay of \xf, for which xenon detectors enriched in \xs\ are less effective. For the two-neutrino decay mode, LZ is predicted to exclude values of the half-life up to 1.7$\times$10$^{24}$ years at 90\% confidence level (CL), and has a three-sigma observation potential of 8.7$\times$10$^{23}$ years, approaching the predictions of nuclear models. For the neutrinoless decay mode LZ, is projected to exclude values of the half-life up to 7.3$\times$10$^{24}$ years at 90\% CL.
\end{abstract}
\maketitle
\section{Introduction}\label{introduction}
Two-neutrino double beta (\tntb) decay is the process by which two neutrons of a given atomic nucleus are converted simultaneously to two protons through the emission of two electrons and two electron antineutrinos,
\begin{equation}
  (Z, A) \rightarrow (Z+2, A)+2e^-+2\bar{\nu}_e.
\end{equation}
This rare process is allowed in the Standard Model of Particle Physics (SM)~\cite{goeppertmayer1935}. Its half-life (\ttn) scales with $Q^{-11}$~\cite{saakyan2013}, where $Q$ is the energy difference between the initial and final nuclear states, and therefore \tntb\ decays with smaller $Q$-values occur at a lower rate. To date, \tntb\ decay has been directly observed in nine nuclides~\cite{pdg2018}, with measured values of \ttn\ up to (2.165$\pm$0.016$_{\mathrm{stat}}\pm$0.059$_{\mathrm{sys}}$)$\times$10$^{21}$ years for \xs~\cite{exo2014}. \tntb\ decay is expected to occur in 26 additional nuclides, with values of \ttn\ that are typically much larger~\cite{haxton1984}.

\xf\ is one nuclide for which \tntb\ decay is expected ($Q=$ 825.8$\pm$0.9 keV~\cite{wang2012}) but not yet confirmed experimentally. The corresponding value of \ttn\ has been calculated in two different nuclear physics models. The predictions for the interacting boson model approximation (IBM-2), that depend on the axial vector coupling parameter ($g_A$), are 3.7$\times$10$^{24}$ and 4.7$\times$10$^{24}$ years for the extreme assumptions $g_A=$ 1.269 and $g_A=$ 1, respectively~\cite{barros2014}. The result from the quasiparticle random-phase approximation (QRPA) is 6.09$\times$10$^{24}$ years~\cite{staudt1990}. Currently the best experimental limit on \ttn\ for \xf\ is 8.7$\times$10$^{20}$ years at 90\% confidence level (CL)~\cite{exo2017}, obtained by EXO-200 using a detector enriched in \xs, with an isotopic abundance of \xf\ of (19.098$\pm$0.0014)\%.

Neutrinoless double beta (\zntb) decay is an alternative decay mode in which no neutrinos are emitted,
\begin{equation}
  (Z, A) \rightarrow (Z+2, A)+2e^-.
\end{equation}
This process is not allowed in the SM and has never been observed experimentally, but if neutrinos are Majorana particles then it would exist in the same nuclides in which \tntb\ decay occurs~\cite{majorana1937, racah1937}. Other extensions of the SM such as supersymmetry or leptoquark theories would also allow for this decay channel~\cite{faessler1998, hirsch1996}. The half-life of the \zntb\ decay (\tzn) is expected to scale with $Q^{-5}$~\cite{saakyan2013}, regardless of the specific short-distance mechanisms. Experiments based on \xs\ have set the strongest constraints on \zntb\ decay to date, excluding values of \tzn\ as large as 1.07$\times$10$^{26}$ years at 90\% CL~\cite{kamlandzen2016}. For \xf, the strongest constraint on \tzn\ has been provided by EXO-200, excluding values up to 1.1$\times$10$^{23}$ years at 90\% CL~\cite{exo2017}.

For light Majorana neutrino exchange, the predicted value of \tzn\ depends on the absolute scale of the neutrino masses as~\cite{doi1981, doi1985}
\begin{equation}
  \frac{1}{T_{1/2}^{0\nu}}=G_{0\nu}|M_{0\nu}|^2\frac{\langle m_{\beta\beta}\rangle^2}{m_e^2},\label{eq_0nu2beta}
\end{equation}
where $G_{0\nu}$ is the phase-space integral of the leptonic contribution to the decay amplitude, $M_{0\nu}$ is the nuclear matrix element of the decay, and $\langle m_{\beta\beta}\rangle$ is the effective Majorana neutrino mass~\cite{dolinski2019}. Eq. \ref{eq_0nu2beta}, combined with the current limits on \zntb\ decay, allows for the exclusion of values of $\langle m_{\beta\beta}\rangle$ down to 0.165 eV at 90\% CL~\cite{kamlandzen2016}.

For other \zntb\ decay mechanisms, such as heavy Majorana neutrino exchange or gluino exchange in supersymmetry models with {\it R}-parity violation ($\slashed{R}$ SUSY), the value of \tzn\ can still be expressed as
\begin{equation}
  \frac{1}{T_{1/2}^{0\nu}}=G_{0\nu}|M_{0\nu}^\mathcal{I}|^2f^\mathcal{I},\label{eq_0nu2beta_gen}
\end{equation}
where $f^\mathcal{I}$ is a factor that contains all the dependence on the new physics parameters~\cite{simkovic2002}. The superindex $\mathcal{I}$ in $M_{0\nu}$ indicates that the value of this quantity depends on the \zntb\ decay mechanism (but not on the value of the new physics parameters). The ratio
\begin{equation}
  R_{0\nu}(\mathrm{N_1}, \mathrm{N_2})\equiv\sqrt{\frac{G_{0\nu}(\mathrm{N_1})\times T_{1/2}^{0\nu}(\mathrm{N_1})}{G_{0\nu}(\mathrm{N_2})\times T_{1/2}^{0\nu}(\mathrm{N_2})}},\label{eq_R_0nu}
\end{equation}
where $\mathrm{N_1}$ and $\mathrm{N_2}$ denote two different nuclides, satisfies
\begin{equation}
  R_{0\nu}(\mathrm{N_1}, \mathrm{N_2})=\frac{|M_{0\nu}^\mathcal{I}(\mathrm{N_2})|}{|M_{0\nu}^\mathcal{I}(\mathrm{N_1})|},
\end{equation}
and is therefore sensitive to the \zntb\ decay mechanism. Pairs of isotopes of the same element such as \xs\ and \xf\ are of particular interest because theoretical uncertainties in the prediction of $R_{0\nu}$ are expected to partially cancel out. The value of $R_{0\nu}(\mathrm{^{136}Xe}, \mathrm{^{134}Xe})$ has been calculated in the framework of renormalized QRPA, obtaining 2.00 for light Majorana neutrino exchange, 3.12 for heavy Majorana neutrino exchange, and 3.03 for gluino exchange in $\slashed{R}$ SUSY~\cite{simkovic2002}.

The energy spectrum of the \tntb\ decay of \xs\ extends up to $Q=$ 2457.83$\pm$0.37 keV~\cite{redshaw2007}, and therefore this process constitutes a background in the search for \tntb\ and \zntb\ decays of \xf. This background is particularly relevant in dedicated xenon detectors designed to search for \zntb\ decay in \xs, such as nEXO~\cite{nexo2018}, NEXT~\cite{next2012}, KamLAND2-Zen~\cite{kamland2zen2020} or PandaX-III~\cite{pandaxiii2017}, as they are enriched in this isotope. The LUX-ZEPLIN dark matter (DM) experiment uses natural xenon instead, for which the isotopic abundances of \xf\ and \xs\ are 10.44\% and 8.87\%, respectively. In this detector, the expected DM signal has properties similar to signals from \tntb\ and \zntb\ decays of \xf, namely rare single-scattering events in the keV to MeV range, that occur at a rate that scales with the size of the active volume. In addition to a relatively low \xs\ content, LZ features a large active mass, very low background levels, accurate fiducialization and good rejection of multiple-scattering events, and hence is expected to be competitive in the search for \tntb\ and \zntb\ decays of \xf.

This article presents the projected sensitivity of LZ to \tntb\ and \zntb\ decays of \xf, based on a profile likelihood ratio (PLR) analysis that uses the energy spectrum of events in an optimized fiducial volume. In Sec. \ref{experiment}, the LZ experiment is reviewed, focusing on the details that are relevant to this study. Background sources are discussed in Sec. \ref{backgrounds}, and the modelling of signal and background is explained in Sec. \ref{models}. The event selection is described in Sec. \ref{selection}. Finally, the sensitivity to the \tntb\ and \zntb\ decays of \xf\ is presented in Sec. \ref{sensitivity}. This section also discusses the potential to constrain the absolute scale of the neutrino masses using the result from the search for the \zntb\ decay of \xf.

\section{The LUX-ZEPLIN experiment}\label{experiment}
\begin{figure*}
  \centerline{\includegraphics[width=0.75\textwidth]{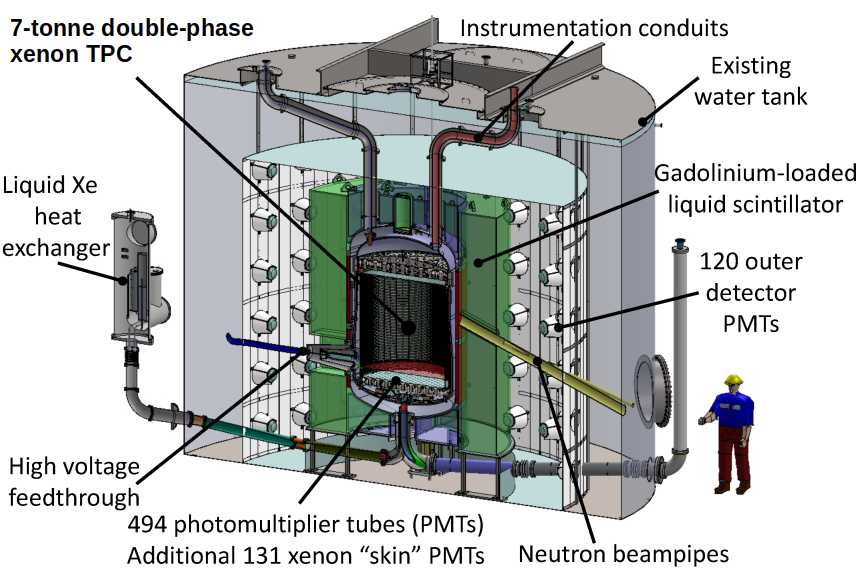}}
  \caption{\label{lz}Cutaway view of the LZ experiment.}
\end{figure*}

The LZ experiment is optimized for the direct detection of DM in the form of weakly-interacting massive particles (WIMPs)~\cite{lz2018}. It is expected to begin operations in 2021. It is located at a depth of 1478 m (4300 m water equivalent) in the Davis Campus at Sanford Underground Research Facility (SURF)~\cite{lesko2015} in Lead, South Dakota, USA. A schematic of the experiment is shown in Fig. \ref{lz}. The complete description of LZ is provided in \cite{lz2017a, lz2019a}, and only details relevant for this study are reviewed here.

The core of the experiment is a cylindrical time-projection chamber (TPC) filled with liquid xenon (LXe), with a small gap at the top filled with gaseous xenon. The TPC is instrumented to measure scintillation and electroluminescence light produced in its volume, with 253 and 241 photomultiplier tubes (PMTs) mounted at the top and the bottom, respectively. The rest of the inner TPC surface is covered by highly reflective polytetrafluoroethylene (PTFE). Four horizontal electrode grids (bottom, cathode, gate and anode) and a series of titanium rings embedded in the PTFE walls provide a nearly uniform electric field inside the TPC directed along its axis. The TPC is contained in a cryostat made of ultrapure titanium~\cite{lz2017b}.

The active volume of the detector is the region of LXe contained between the cathode and gate grids. Both the diameter and the height of the active LXe volume are 1.46 m, resulting in a mass of 7 tonnes, which corresponds to 741 kg of \xf. When an incoming particle interacts in the active volume, the kinetic energy of the recoiling nucleus or electron is transferred to the medium, generating a detectable prompt scintillation light (S1) and ionization electrons. The ionization electrons are drifted towards the gaseous xenon phase at the top of the TPC by the applied electric field, where they are extracted from the LXe and then accelerated in the gaseous phase using a higher electric field applied between the gate and anode electrodes, emitting detectable electroluminescence light (S2). The signals measured by LZ consist of the prompt S1 plus the delayed S2~\cite{chepel2013}. The delay between the S1 and S2 signals is used to reconstruct the depth of particle interactions, while the distribution of collected light over the photomultipliers in the top array is used to reconstruct the radial position.

The interval of energy to search for the \tntb\ and \zntb\ decays of \xf\ extends up to nearly 1 MeV, while the design of the LZ detector is optimized to measure energy depositions below 100 keV. For this reason it is expected that the events in the region of interest of this analysis will be affected by PMT saturation. As in the search for the \zntb\ decay of \xs, the energy measurement can avoid these effects by using only the S2 signal provided by the bottom PMT array, which are not expected to saturate, even at the highest energies relevant to this analysis~\cite{lz2020a}.

The TPC is surrounded by two active vetoes, used to discriminate background events with multiple interaction vertices, and a water tank for passive shielding. The first active veto, called the \textit{xenon skin}, consists of an instrumented layer of 2 tonnes of additional LXe filling the lateral and bottom spaces between the TPC and the inner cryostat vessel. The xenon skin is optically isolated from the TPC volume, and observed by 98 and 38 independent PMTs mounted at the top of the lateral space and at the bottom of the TPC, respectively. The objective of the xenon skin is to identify multiple-scattering events by measuring scintillation light in coincidence with events in the TPC. The second active veto, called the \textit{outer detector} (OD), consists of a nearly-hermetic layer of 17 tonnes of gadolinium-loaded liquid scintillator (GdLS) surrounding the cryostat, observed by 120 PMTs mounted in the water tank. The main objective of the OD is to identify multiple-scattering neutrons by measuring the $\sim$8 MeV cascade of gamma rays from their capture on gadolinium in coincidence with events in the TPC. However, the OD can also detect high energy depositions from other particles such as external photons. The entire setup is installed inside a nearly-hermetic layer of 228 tonnes of ultrapure water. The objective of this passive shielding is to suppress the radiation from outside the experiment and from the OD PMTs. It is also used as an active muon veto.

LZ will conduct a comprehensive calibration program using a variety of radiation sources~\cite{lz2019a}. In particular, the energy resolution of the detector at the $Q$-value of \xf\ will be assessed using an external source of $^{54}$Mn, which emits a 834.9 keV gamma ray after decaying by electron capture to the 2$^+$excited level of $^{54}$Cr. Neutron calibrations will produce short-lived isotopes in the xenon target, such as $^{127}$Xe, $^{131m}$Xe or $^{133}$Xe, which provide characteristic decay lines that can be used to determine the energy resolution around the maximum of the \tntb\ decay spectrum of \xf\ ($\sim$200 keV). The use of external sources of $^{22}$Na and $^{228}$Th can also be used for that purpose. Additional calibration lines will also be available from the metastable $^{85m}$Kr or $^{131m}$Xe isotopes that will be regularly injected into the xenon to study the position dependence of the detector response.
\phantom{\ref{systematics}}

\begin{table*}
  \centering
  \renewcommand{\arraystretch}{1.25}
  \begin{tabular}{lr}
    \hline
    \hline
    Background & \multicolumn{1}{c}{Uncertainty} \\
    & \multicolumn{1}{c}{[\%]} \\
    \hline
    \tntb\ decay of \xs & 3 \\
    Solar neutrinos & 2 \\
    Beta decay of \ke & 20 \\
    Decay chain of \rt & 10 \\
    Decay chain of \rz & 10 \\
    Gamma decays outside LXe & 20 \\
    \hline
    \hline
  \end{tabular}
  \caption{Systematic uncertainties assumed in the normalization of the background sources discussed in Sec. \ref{backgrounds}, included in the PLR as Gaussian nuisance parameters.}
  \label{systematics}
\end{table*}

\section{Background sources in the search for \tntb\ and \zntb}\label{backgrounds}
The backgrounds assessment is similar to that for other sensitivity studies of LZ, such as for weakly-interacting massive particles (WIMPs)~\cite{lz2018}, the \zntb\ decay of \xs~\cite{lz2020a}, or new physics via low-energy electron recoils~\cite{lz2021}. Six background contributions are found to contribute a significant number of events to the region of interest of the analysis considered here:
\begin{enumerate}
\item \tntb\ decay of \xs\ in LXe: The isotopic abundance of \xs\ in natural xenon is 8.9\%, which implies 646 kg of this isotope in the active volume. The half-life of this decay is (2.165$\pm$0.016$_{\mathrm{stat}}\pm$0.059$_{\mathrm{sys}}$)$\times$10$^{21}$ years~\cite{exo2014}, and therefore approximately 3.56 million events are expected for a live time of 1000 days. The energy spectrum extends up to $Q=$ 2457.83$\pm$0.37 keV~\cite{redshaw2007}.
\item Gamma rays from radioactivity in experiment components and cavern walls: This radiation is caused by de-excitation of daughter nuclei after alpha or beta decays ocurring in materials surrounding the LXe volume. Radioactive contaminants include nuclides of the $^{238}$U and $^{232}$Th chains, $^{40}$K, and $^{60}$Co. The activity of experiment components has been assessed by means of an intensive screening program~\cite{lz2020b}, while that of cavern walls has been determined from an \textit{in-situ} measurement of the radiation fluxes~\cite{lz2019c}. Based on the background model described in Sec. \ref{models}, the sources providing the dominant contributions are the rings that shape the electric field, the cryostat vessels, and the cavern walls.
\item Decay chain of \rt\ dissolved in LXe: \rt\ enters LXe by emanation from detector materials and dust, which are estimated to contribute approximately 80\% and 20\% of the total, respectively~\cite{lz2018}. The background is dominated by the beta decay of $^{214}$Pb ($Q=$ 1019 keV). The resulting $^{214}$Bi nucleus is produced directly in the ground state with 9.2\% probability, and in this case, a single electron recoil is observed. Otherwise, the beta electron is accompanied by gamma rays from the de-excitation of the $^{214}$Bi nucleus, although a single electron recoil can still be observed if such photons escape from the active volume. The subsequent beta decay of $^{214}$Bi is excluded because it is typically detected in coincidence with the alpha decay of its daughter ($^{214}$Po), which has a half-life of 162 $\mu$s, leading to a 99.99\% rejection of $^{214}$Bi beta decays occurring in the active region~\cite{lz2020a}. Finally, long-lived nuclides are assumed to be extracted from the bulk of the active region before they decay~\cite{exo2015}. For this reason, the beta decays of $^{210}$Pb (that has a half-life of 22.6 years) and its progeny are excluded. The activity of \rt\ is assumed to be equal to the LZ design requirement of 2 $\mu$Bq/kg~\cite{lz2017a}.
\item Decay chain of \rz\ in LXe: Similar to \rt, \rz\ enters LXe by emanation from detector materials and dust. The dominant process is the beta decay of $^{212}$Pb ($Q=$ 570 keV), that proceeds directly to the ground state of $^{212}$Bi with 13.3\% probability. In this case the decay is observed as a single electron recoil, without any accompanying gamma rays from the de-excitation of the $^{212}$Bi nucleus. The beta decay of $^{212}$Bi can be rejected with virtually 100\% efficiency because it is detected in coincidence with the alpha decay of its daughter, $^{212}$Po, which has a half-life of 0.299 $\mu$s, and for this reason it is excluded from the background model. The activity of \rz\ is assumed to be 5\% that of \rt, based on the ratio seen in LUX~\cite{lux2016}, and therefore equal to 0.1 $\mu$Bq/kg.
\item Beta decay of \ke\ dissolved in LXe: This decay proceeds directly to the ground state of $^{85}$Rb with 99.56\% probability, and thus the majority of events consist of a single electron recoil with no accompanying gamma rays (``naked'' beta). The resulting energy spectrum is similar to that of the \tntb\ decay of \xf, with $Q=$ 698.4 keV, and therefore it could have a severe impact on the sensitivity. This fact is further discussed in Secs. \ref{sensitivity} and \ref{conclusions}. The concentration of natural krypton dilluted in the xenon volume is assumed to be 0.3 ppt g/g $^{nat}$Kr/Xe based on the design requirement~\cite{lz2017a}. This concentration will be achieved by chromatographic separation on charcoal before the start of physics data-taking~\cite{lux2018a}, and will be verified \textit{in-situ} using mass spectrometry. In addition, the amount of $^{85}$Kr can also be determined once data-taking begins by measuring the rate of the subdominant decay branch of this isotope (0.44\% branching fraction), which involves detecting a beta decay in coincidence with the subsequent gamma decay (1.015 $\mu$s half-life). The isotopic abundance of \ke\ in natural krypton is assumed to be 2$\times$10$^{-11}$~\cite{collon2004, du2003}.
\item Electron recoils from solar neutrino interactions: The spectrum is assumed to be dominated by the $pp$ and $^{7}$Be neutrinos of the $pp$ chain, and the $^{13}$N neutrinos of the CNO cycle. Other contributions are expected to be subdominant with respect to the rest of backgrounds considered, and hence are not included.
\end{enumerate}

Liquid xenon flowing through the purification system will not be shielded by the outer detector and the water tank, and therefore it will experience an increased activation rate from environmental thermal neutrons. Among the isotopes resulting from neutron activation, three decay via beta emission and are therefore potential background sources for this analysis: $^{133}$Xe ($Q=$ 427 keV), $^{135}$Xe ($Q=$ 1151 keV), and $^{137}$Xe ($Q=$ 4162 keV). The half life of these decays is 5.24 days, 9.14 hours and 3.82 minutes, respectively. After completing the purification cycle, these radioactive isotopes may reach the cryostat and decay in the active volume of the TPC. The corresponding energy deposition would be detected as a single interaction if the beta decay proceeds directly to the ground state of the daughter nuclide, or it is accompanied by conversion electrons only. The number of events resulting from these processes was estimated to be subdominant with respect to any of the sources listed above, and to contribute less than 1\% to the total background of both the \tntb\ and \zntb\ decays. Consequently, this contribution is not included in the background model.

\begin{figure*}
  \includegraphics[width=0.48\textwidth]{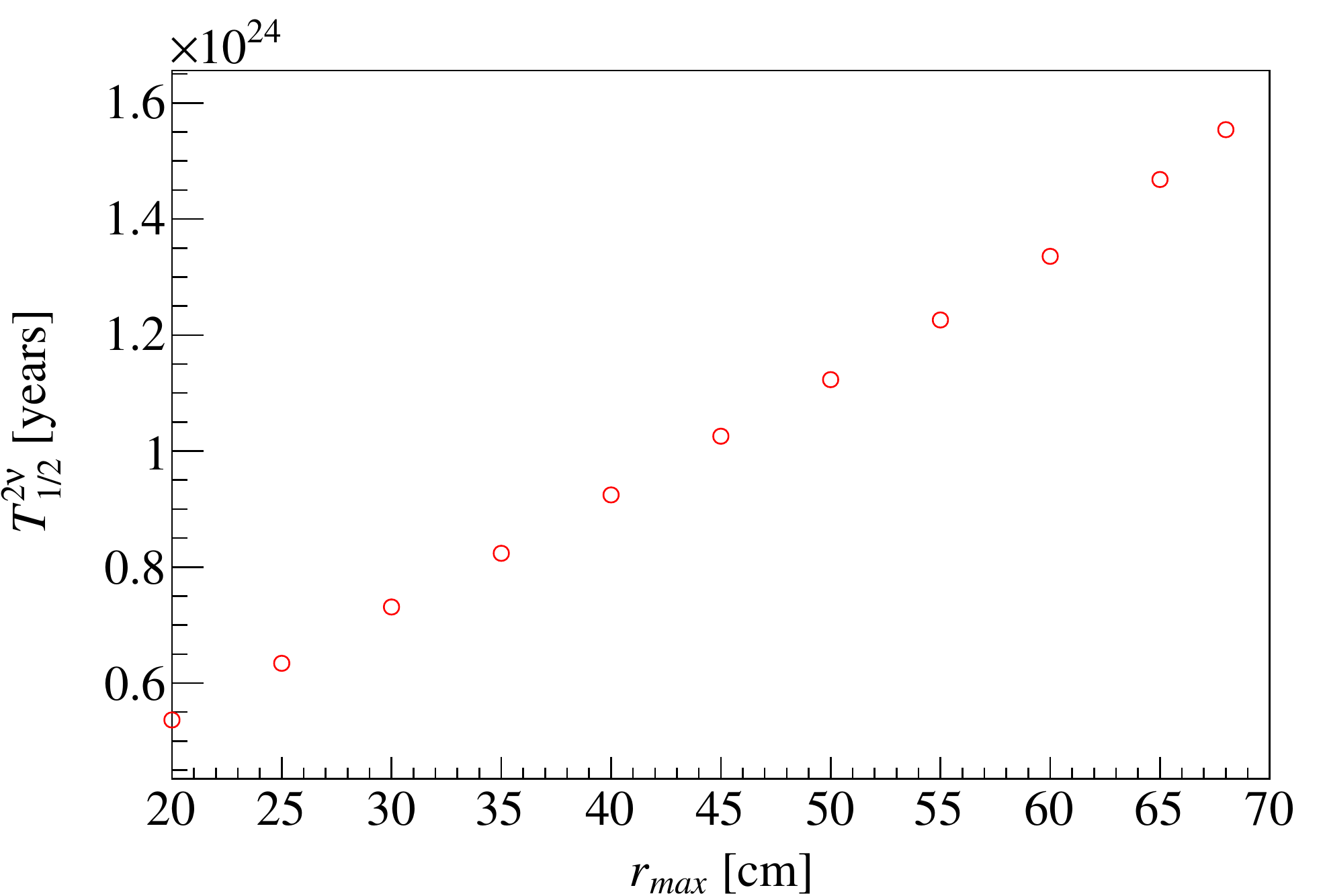}\hfill
  \includegraphics[width=0.48\textwidth]{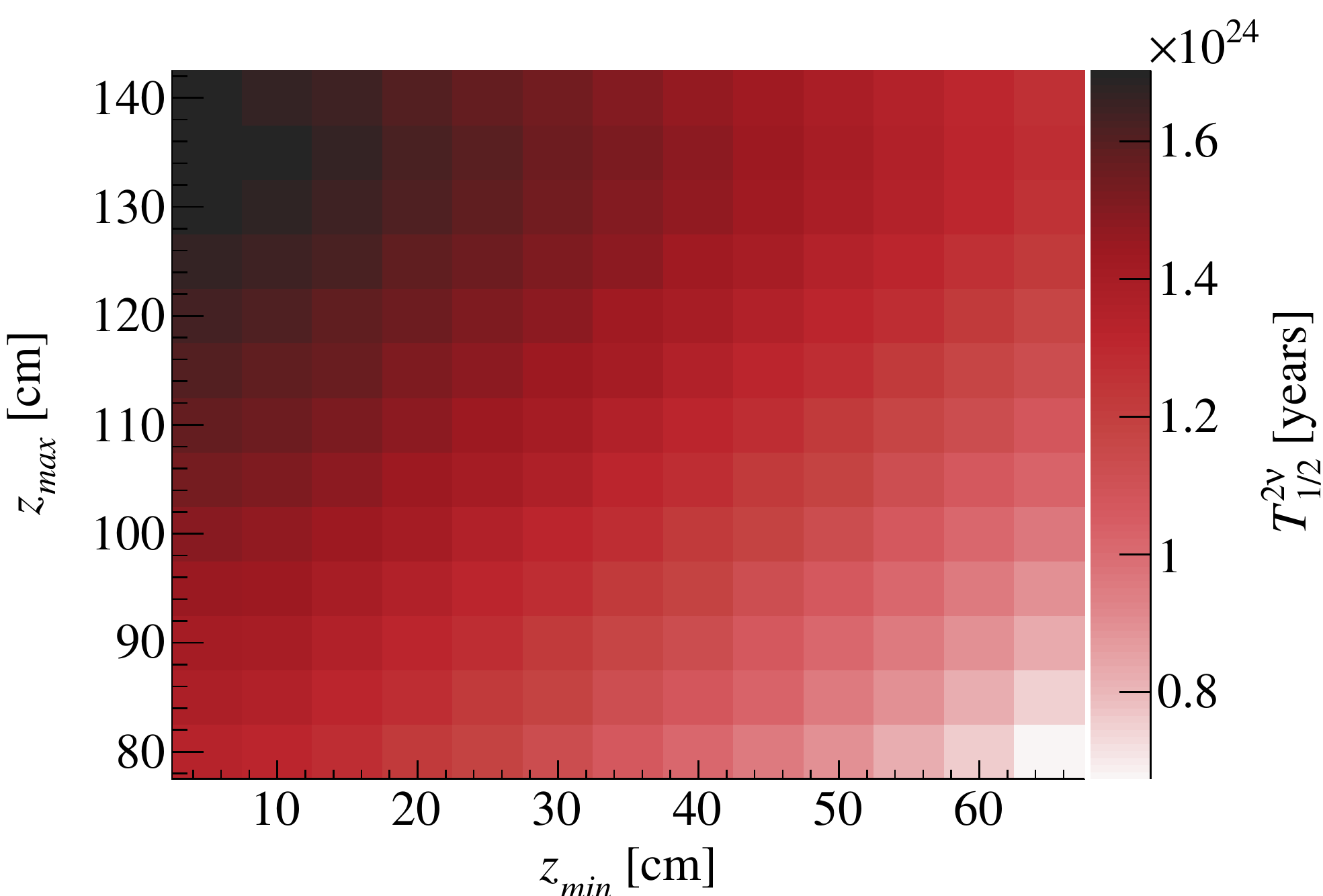}
  \includegraphics[width=0.48\textwidth]{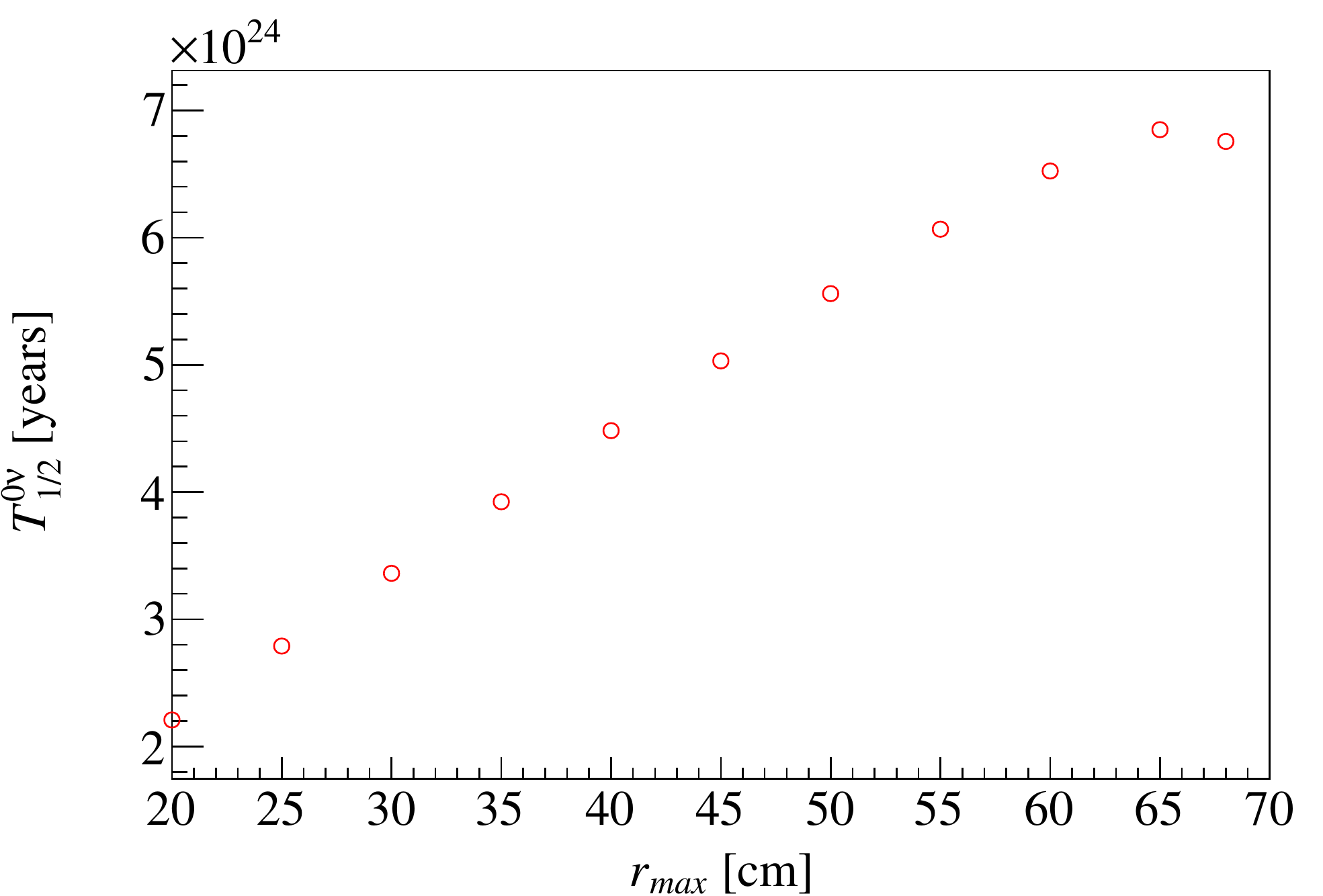}\hfill
  \includegraphics[width=0.48\textwidth]{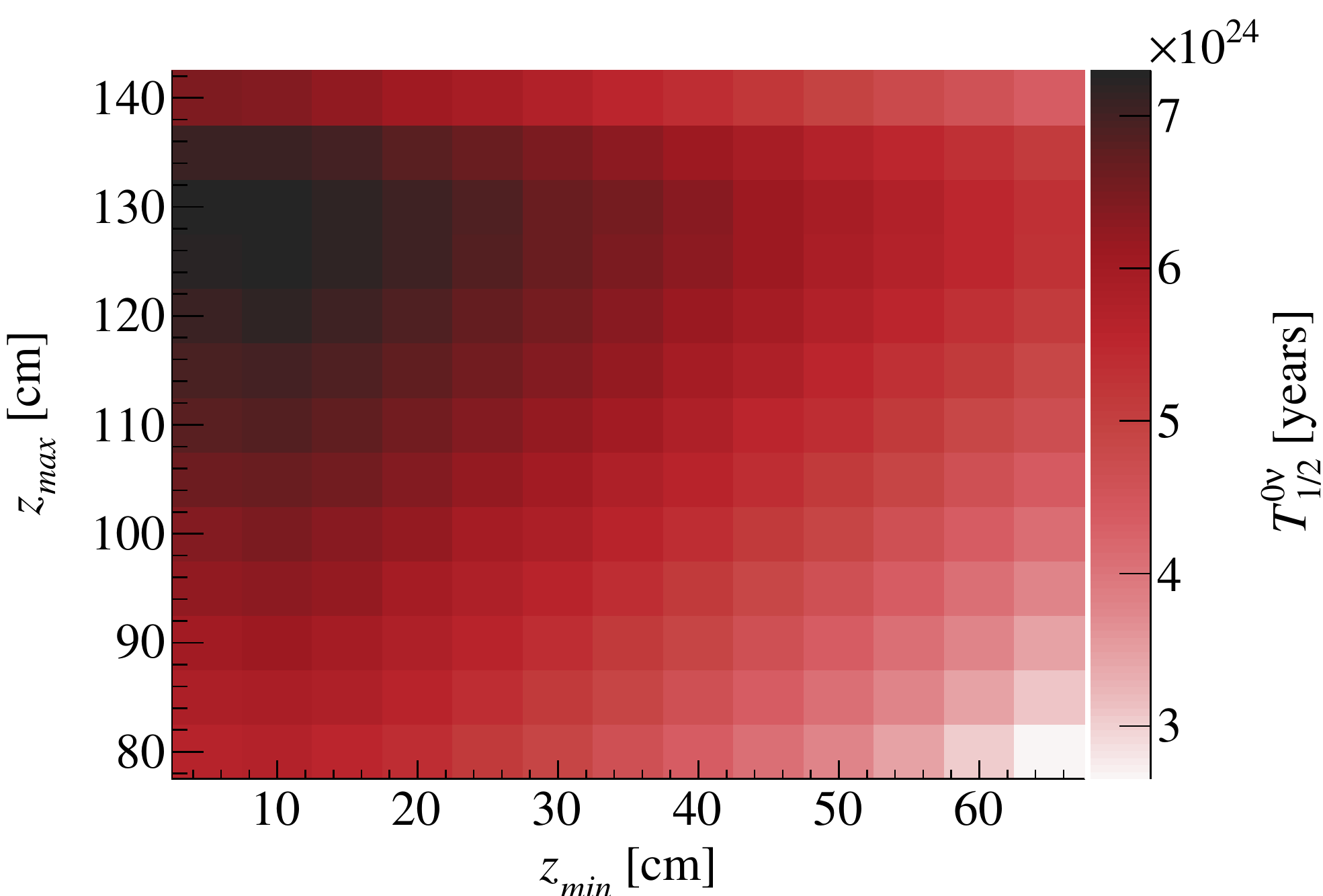}
  \caption{Top left: sensitivity to \ttn\ of \xf\ as a function of $r$, for $z_{min}=$ 25 cm and $z_{max}=$ 125 cm. Top right: sensitivity to \ttn\ of \xf\ as a function of $z_{min}$ and $z_{max}$, for $r=$ 68.8 cm. Bottom left: sensitivity to \tzn\ of \xf\ as a function of $r$, with $z_{min}=$ 25 cm and $z_{max}=$ 125 cm. Bottom right: sensitivity to \tzn\ of \xf\ as a function of $z_{min}$ and $z_{max}$, for $r=$ 65 cm.}
  \label{scan}
\end{figure*}

All the backgrounds discussed above consist of electron recoils. Neutron backgrounds, relevant to the WIMP sensitivity study of LZ~\cite{lz2018}, consist of nuclear recoils instead. While the total rate for electron recoil backgrounds is $\mathcal{O}$(10$^{-4}$) through $\mathcal{O}$(10$^{-3}$) counts kg$^{-1}$ day$^{-1}$ keV$^{-1}$, the total rate for neutron backgrounds lies below 10$^{-8}$ counts kg$^{-1}$ day$^{-1}$ keV$^{-1}$ in the interval of energies of interest for the current analysis~\cite{lz2018}. In addition, this background contribution can be further suppressed thanks to the discrimination between electron recoils and nuclear recoils in LXe. Based on these facts, the neutron background is not included.
\phantom{\ref{scan}}
  
\section{Signal and background models}\label{models}
The \tntb\ and \zntb\ decay signals, and the background contributions 1 and 6 in Sec. \ref{backgrounds}, consist of single-scattering events distributed uniformly in LXe. Their energy spectra are built using existing numerical data or analytical functions as described below. For the remaining background sources, the energy spectra are built from Monte Carlo (MC) simulations generated with BACCARAT~\cite{lz2020c}, a software package based on GEANT4~\cite{geant2003, geant2016} (version 9.5.p02) that provides a generic framework to simulate the response of noble gas detectors. Gaussian smearing is applied to all energy spectra in order to account for detector resolution effects. The energy resolution function has been calculated using the Noble Element Simulation Technique (NEST) software~\cite{szydagis2011, nest2018}, assuming the projected detector performance considered in previous sensitivity calculations~\cite{lz2018, lz2020a, lz2021}. The energy resolution at the maximum of the \tntb\ decay spectrum of \xf\ ($\sim$200 keV) is approximately 2.6\%. At the $Q$-value of these decays (825.8 keV) the resolution is 1.64\%.

The energy spectrum of the \tntb\ decay of \xf\ is built using numerical data provided by the Nuclear Theory group at Yale University~\cite{yale, kotila2012}. For the \zntb\ decay of \xf, the energy spectrum is modelled as a single line at $Q=$ 825.8 keV. These spectra account for decays to the ground state of the daughter nucleus ($^{134}$Ba), and also for decays to the 2$^+$ state in which the accompanying gamma ray (605 keV) is completely measured in the detector~\cite{dama2002}. The effect of not fully detecting the accompanying gamma ray in the latter case is not modelled due to the absence of a prediction of the relative branching fraction of \xf\ decays to the 2$^+$ state of $^{134}$Ba.

Although both \tntb\ and \zntb\ decays of \xf\ are assumed to consist of single scatters distributed uniformly over LXe, there is a small probability of having signal events in which bremsstrahlung photons create additional scatters that can be spatially resolved. This probability has been calculated using a dedicated MC simulation, obtaining (2.13$\pm$0.06)\% for the \zntb\ decay of \xf. This fraction is expected to be smaller for the \tntb\ decay of \xf, because the total kinetic energy of the emitted electrons is less than $Q$. Based on these results, the contribution of multiple-scattering signal events is neglected.

\begin{figure*}
  \includegraphics[width=0.48\textwidth]{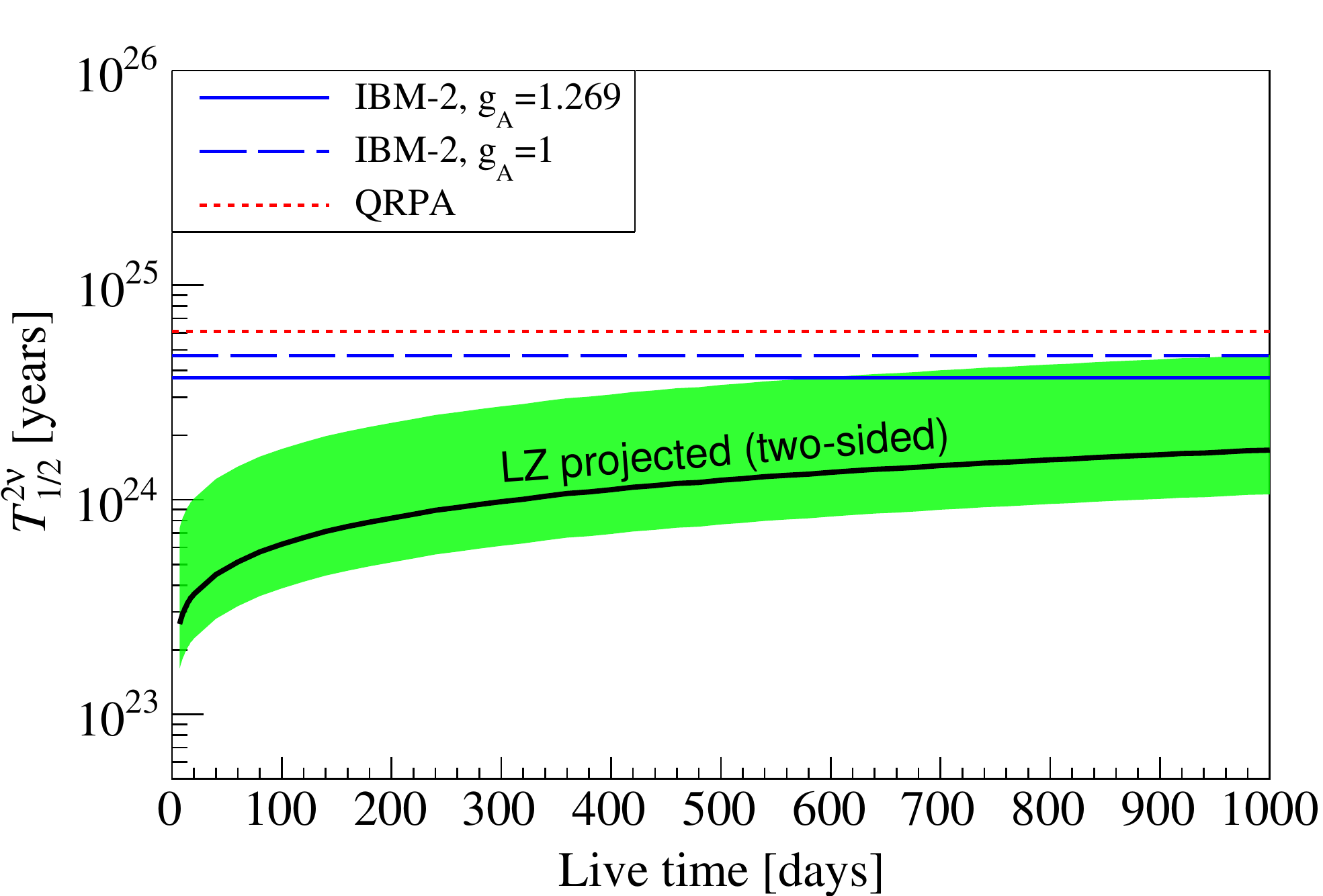}\hfill
  \includegraphics[width=0.48\textwidth]{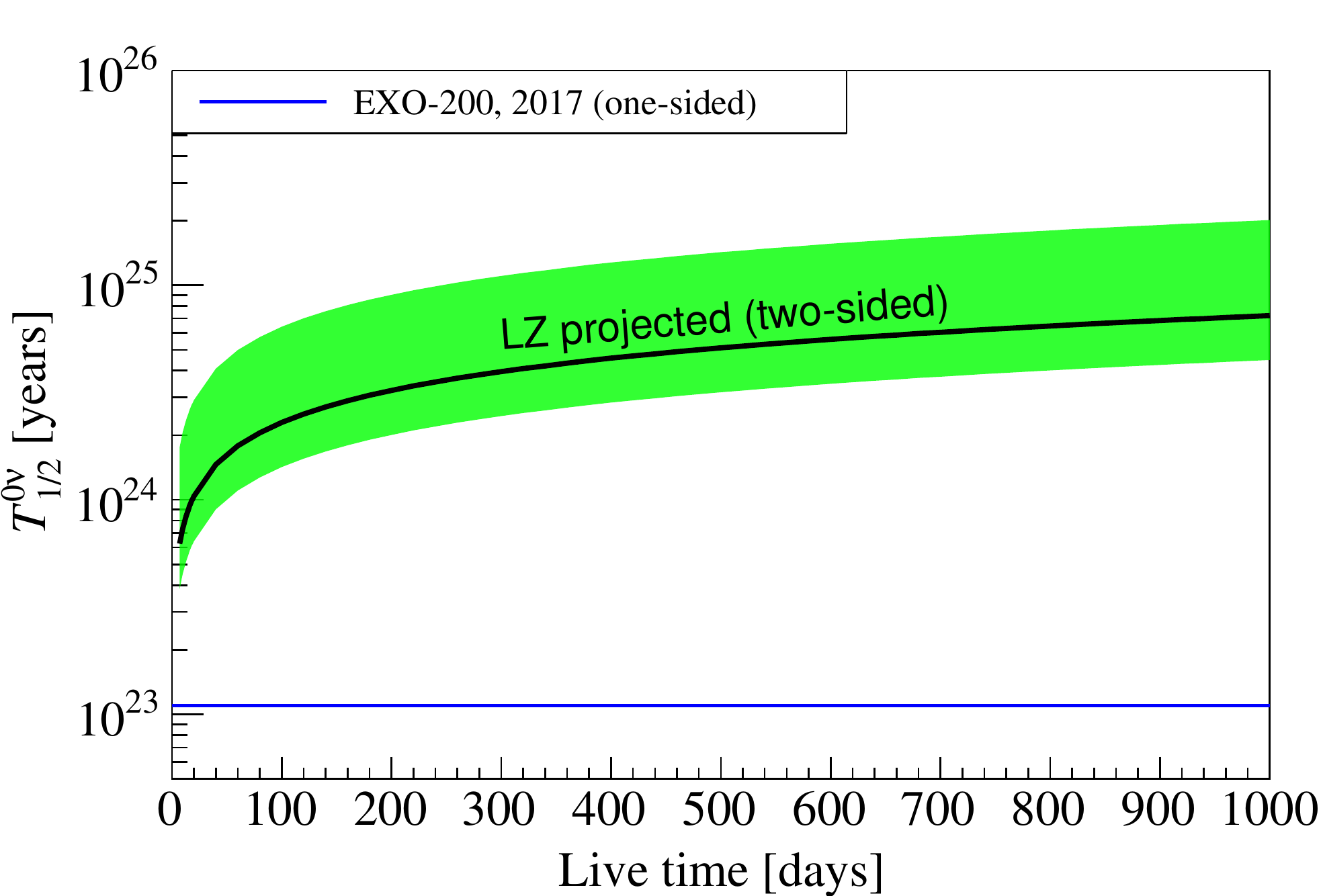}
  \caption{Left: sensitivity to the \tntb\ decay of \xf\ as a function of the live time of the detector (black), along with the respective statistical uncertainty at 1$\sigma$ (green band). The horizontal lines show the predictions for IBM-2~\cite{barros2014}, assuming $g_A=1.269$ (continuous blue) and $g_A=1$ (dashed blue), and QRPA~\cite{staudt1990} (dotted red). The current best limit, set by EXO-200~\cite{exo2017}, is 8.7$\times$10$^{20}$ years at 90\% CL, and therefore lies below the minimum of the vertical axis. Right: sensitivity to the \zntb\ decay of \xf\ as a function of the live time of the detector (black), along with the respective statistical uncertainty at 1$\sigma$ (green band). The horizontal line (blue) shows the current best limit, set by EXO-200~\cite{exo2017}.}
  \label{sens}
\end{figure*}

The energy spectrum of the \tntb\ decay of \xs\ is also built using numerical data provided by the Nuclear Theory group at Yale University~\cite{yale, kotila2012}. This spectrum is normalized to the event rate that corresponds to the expected activity of \xs, assuming \ttn\ equal to 2.165$\times$10$^{21}$ years~\cite{exo2014}.
\phantom{\ref{sens}}

For the solar neutrino background, the energy spectrum is built using an analytical function~\cite{bahcall1987} modified to include the effect of the electron binding energy in xenon atoms~\cite{chen2017}. The spectrum obtained from such function is already normalized to the correct event rate per unit mass.

All the energy spectra obtained from MC simulations are normalized to the event rate that corresponds to the respective expected activities, already discussed in Sec. \ref{backgrounds}.

\section{Event selection}\label{selection}
Signal events are selected by requiring single scatters within an energy window between 5 keV and 1 MeV, in order to contain the full spectrum of both \tntb\ and \zntb\ decays of \xf. Multiple scatters in LXe are rejected using standard criteria common to other analyses of LZ~\cite{lz2018}, that require the energy-weighted dispersion of interaction positions to be below 3 and 0.2 cm along the radial and vertical directions, respectively. These cut values are based on the spatial resolution observed by the LUX experiment~\cite{lux2018b, solovov2012}. Multiple scatters involving the active vetoes of the experiment are rejected by requiring that both the xenon skin and the OD measure an energy deposit below 100 keV within a time window of 100 $\mu$s before and after the primary interaction.

For each decay channel (\tntb\ and \zntb) a fiducial volume (FV) is defined as a cylinder with a given radius $r$, minimum height $z_{min}$ and maximum height $z_{max}$, contained inside the active region. $r$, $z_{min}$ and $z_{max}$ are optimized for each of the two decay channels separately in order to maximize the sensitivity of the analysis. The optimization procedure and the resulting FVs are explained in Sec. \ref{sensitivity}. By convention, $z_{min}$ and $z_{max}$ are measured from the bottom of the sensitive LXe volume. For MC-simulated background samples, the FV cut is applied by only accepting events for which the true position is contained within $r$, $z_{min}$ and $z_{max}$. For event populations that consist of single scatters distributed uniformly over LXe, the energy spectra is scaled by the ratio of the FV to the total LXe volume.

\section{Sensitivity projections}\label{sensitivity}
The projected sensitivity of LZ to \tntb\ and \zntb\ decays of \xf\ is calculated assuming an experimental live time of 1000 days. For each decay, the sensitivity is defined as the median of the lower limits on the half-life, set at 90\% CL, that would be obtained by successive experiments if the background-only hypothesis were true. The calculations use the PLR method with the asymptotic two-sided test statistic~\cite{cowan2011}, that provides a nearly-optimal performance and allows the inclusion of systematic uncertainties. The sensitivity is found by performing a frequentist hypothesis test inversion, using the RooStats package~\cite{moneta2010}. In addition, an analogous calculation is carried out to determine the maximum value of \ttn\ that could be observed at three-sigma level, also using the asymptotic two-sided test statistic.
\phantom{\ref{counts}}

\begin{table*}
  \centering
  \renewcommand{\arraystretch}{1.25}
  \begin{tabular}{lrr}
    \hline
    \hline
    Contribution & \multicolumn{1}{c}{Counts} & \multicolumn{1}{c}{Counts} \\
    & \multicolumn{1}{c}{\tntb\ selection} & \multicolumn{1}{c}{\zntb\ selection} \\
    \hline
    \tntb\ decay of \xs & 80\,100 & 119\,000 \\
    Solar neutrinos & 4\,800 & 0 \\
    Beta decay of \ke & 22\,200 & 0 \\
    Decay chain of \rt & 22\,800 & 17\,000 \\
    Decay chain of \rz & 6\,480 & 0 \\
    Gamma rays & 38\,200 & 29\,700 \\
    \hline
    Total background & 175\,000 & 166\,000 \\
    Signal & 1\,560 & 560 \\
    \hline
    \hline
  \end{tabular}
  \caption{Event counts for the background categories discussed in Sec. \ref{backgrounds} for a live time of 1000 days, in the sensitive region of each analysis, using the event selections explained in Sec. \ref{selection} along with the optimized FV requirements found in Sec. \ref{sensitivity}. The sensitive regions, defined in Sec. \ref{sensitivity}, are the interval between 5 and 250 keV for the \tntb\ decay, and the window of 40 keV around $Q=$ 825.8 keV for the \zntb\ decay. The event counts are rounded to the precision set by the statistical uncertainty.}
  \label{counts}
\end{table*}

The PLR developed for this work uses only information from the energy spectrum. The total background spectrum is built by adding the six contributions discussed in Sec. \ref{backgrounds} and scaling the resulting spectrum by the live time. The systematic uncertainty in the normalization of these contributions is accounted for by Gaussian nuisance parameters, following closely the procedure developed in the WIMP sensitivity study of LZ~\cite{lz2018} (see Table \ref{systematics}). The uncertainty for the \tntb\ decay of \xs\ is taken from the latest measurement of its \ttn~\cite{exo2014}, while that for the solar neutrinos is taken from their flux measurements~\cite{bahcall2005}. The remaining uncertainties are those estimated for the respective \textit{in-situ} background measurements that will be carried out in LZ, based on the performance of such studies in LUX~\cite{lux2015a, lux2015b}.

The sensitivity defined above serves as the figure of merit to optimize the FV described in Sec. \ref{selection}. This optimization is carried out separately for each decay channel of \xf\ by finding the maximum sensitivity over a range of values of $r$, $z_{min}$ and $z_{max}$, using a two-step scanning procedure (see Fig. \ref{scan}). First, $r$ is scanned while $z_{min}$ and $z_{max}$ are fixed to some initial values. Second, $z_{min}$ and $z_{max}$ are scanned simultaneously while $r$ is fixed to the value providing the maximum sensitivity in the previous iteration. The values resulting from the FV optimization are $r=$ 68.8 cm, $z_{min}=$ 5 cm and $z_{max}=$ 135 cm for \tntb\ decay, and $r=$ 65 cm, $z_{min}=$ 10 cm and $z_{max}=$ 130 cm for \zntb\ decay. The resulting FV contains 5.44 and 4.59 tonnes of LXe, respectively. The robustness of each optimization result is checked by redoing the scan over $z_{min}$ and $z_{max}$ for the values of $r$ adjacent to the optimal one, and confirming that the sensitivity does not improve.

The sensitivity is found to be 1.7$\times$10$^{24}$ years for \ttn\ and 7.3$\times$10$^{24}$ years for \tzn\ after 1000 live days. Therefore, it will be possible to reach the domain of the \ttn\ predictions from the IBM-2 and QRPA models (see Fig. \ref{sens}), while the lower limit for \tzn\ will improve by almost two orders of magnitude with respect to the existing experimental constraints. In addition, it is found that the three-sigma observation potential of LZ to \ttn\ is 8.7$\times$10$^{23}$ years, for the optimal values of $r$, $z_{min}$ and $z_{max}$ obtained above. If the asymptotic one-sided test statistic is used instead, to allow a direct comparison with previous results~\cite{exo2017}, the exclusion limits change to 2.2$\times$10$^{24}$ and 9.4$\times$10$^{24}$ years for the \tntb\ and \zntb\ decays, respectively.

\begin{figure*}
  \includegraphics[width=0.48\textwidth]{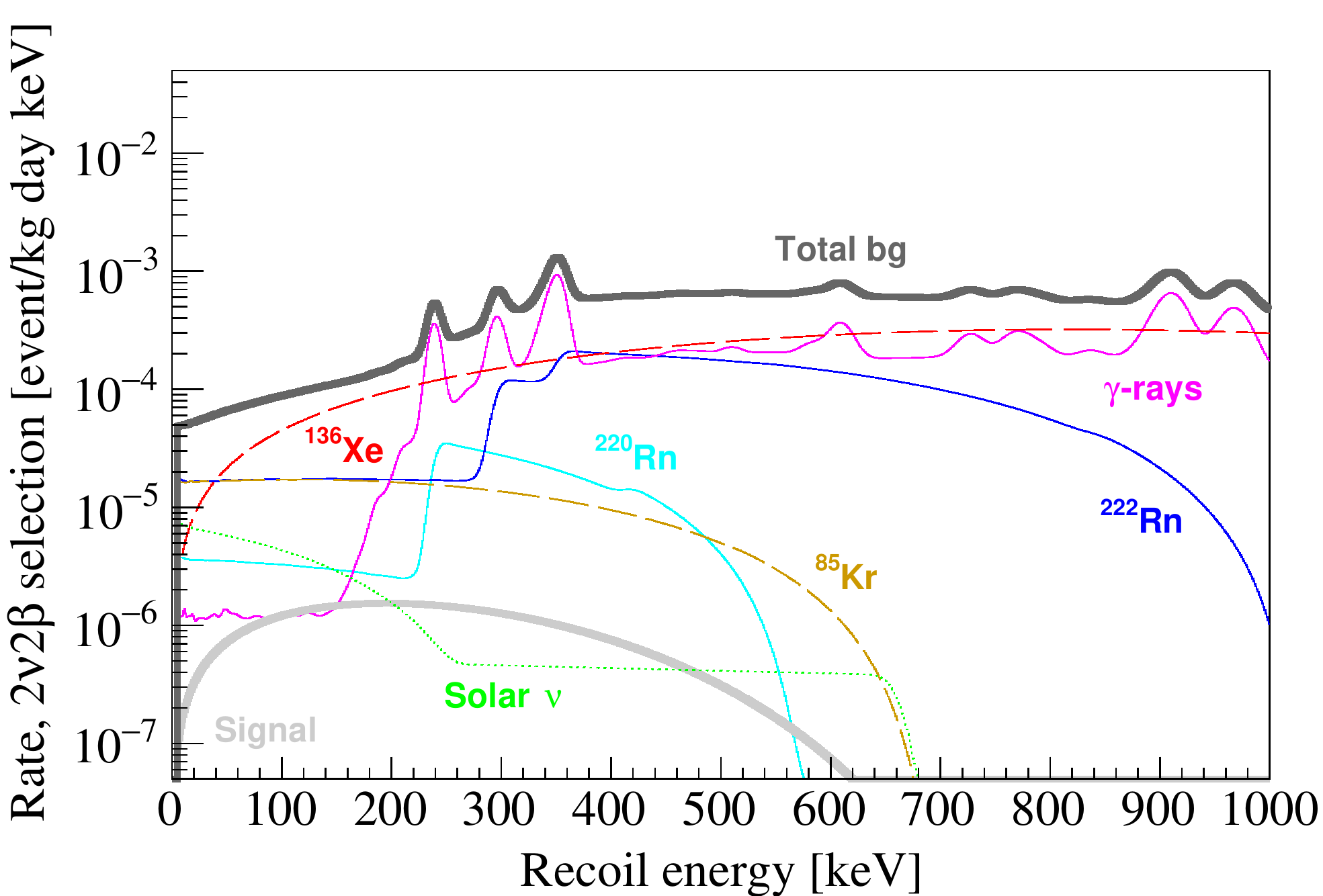}
  \includegraphics[width=0.48\textwidth]{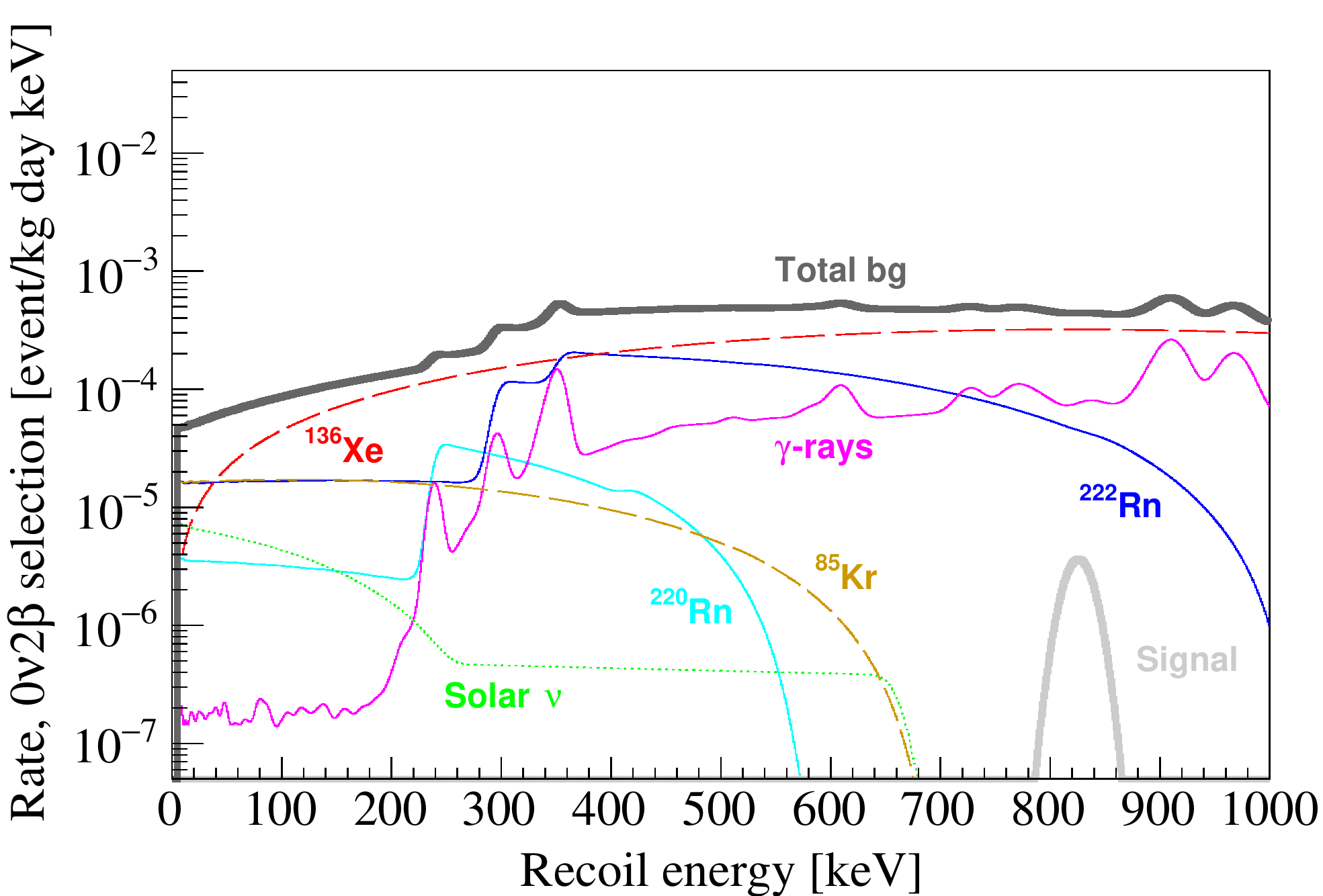}
  \caption{Energy spectra of the \tntb\ (left) and \zntb\ (right) decays of \xf, along with those of the background categories described in Sec. \ref{backgrounds}. In each case the signal assumes the respective 90\% CL half-life obtained in this study. The spectra were obtained using the event selection described in Sec. \ref{selection} along with the respective optimal FV found in Sec. \ref{sensitivity}. The curves show the signal (continuous light gray) and the total background (continuous dark gray), along with the partial contributions from the \tntb\ decay of \xs\ (dashed red), solar neutrinos (dotted green), the beta decay of \ke\ (dashed orange), the decay chains of \rt\ (continuous blue) and \rz\ (continuous cyan), and gamma rays from the contamination in the detector components and the cavern walls (magenta).}
  \label{spectra}
\end{figure*}

Fig. \ref{spectra} shows the energy spectra of signal and background, using the optimal values of $r$, $z_{min}$ and $z_{max}$, and assuming the sensitivity values of \ttn\ and \tzn\ for \xf. For each analysis a sensitive region (SR) can be defined as the energy interval that maximizes the statistical significance $S/\sqrt{B}$, where $S$ and $B$ are the total number of signal and background events, respectively. The SR for the \tntb\ decay search is found to be the interval from 5 keV (low energy limit of the analysis) to 250 keV, while that for the \zntb\ decay search is a 40 keV window around $Q=$ 825.8 keV. The total event counts in each SR are summarized in Table \ref{counts}.
  
\begin{figure*}
  \includegraphics[width=0.48\textwidth]{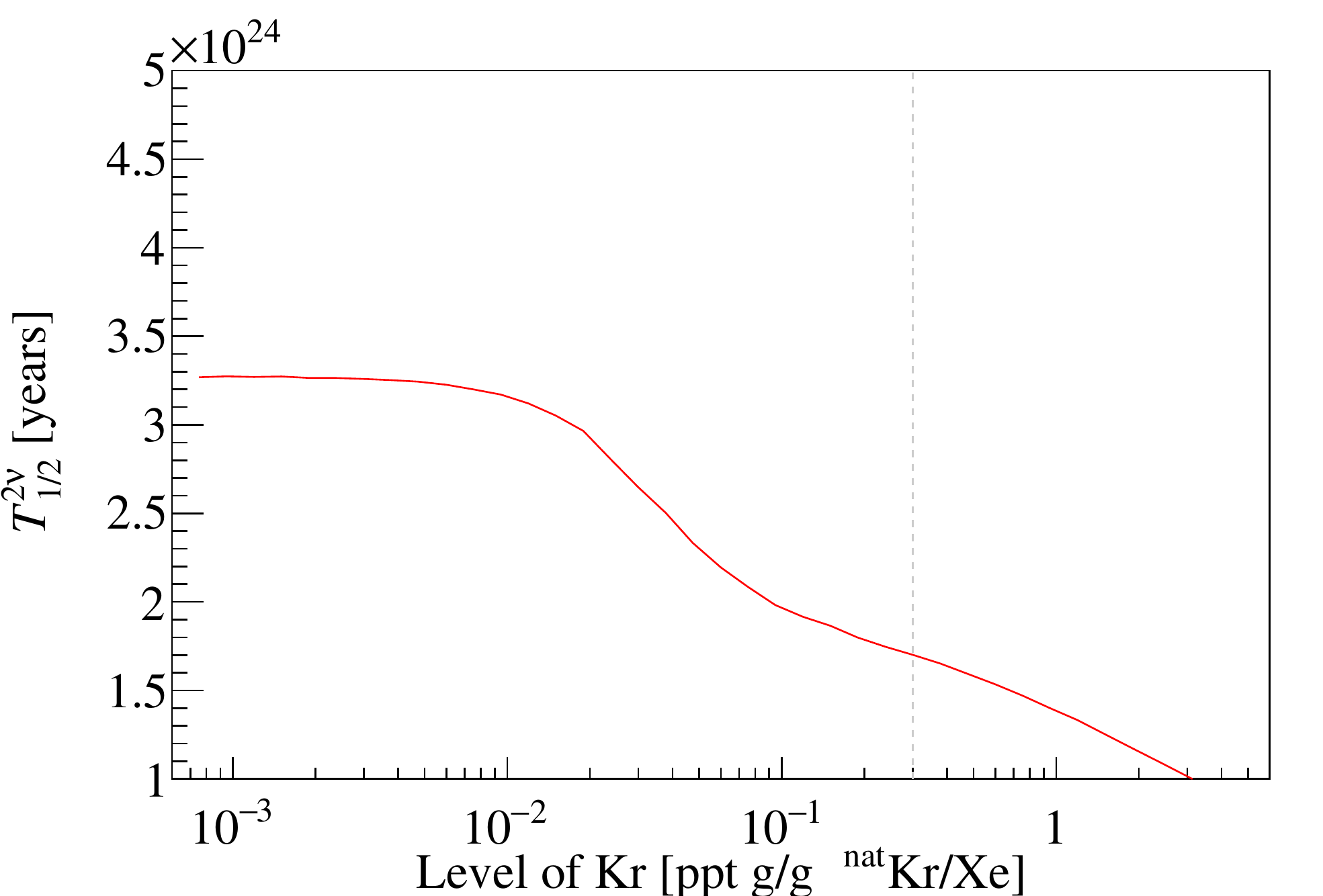}
  \includegraphics[width=0.48\textwidth]{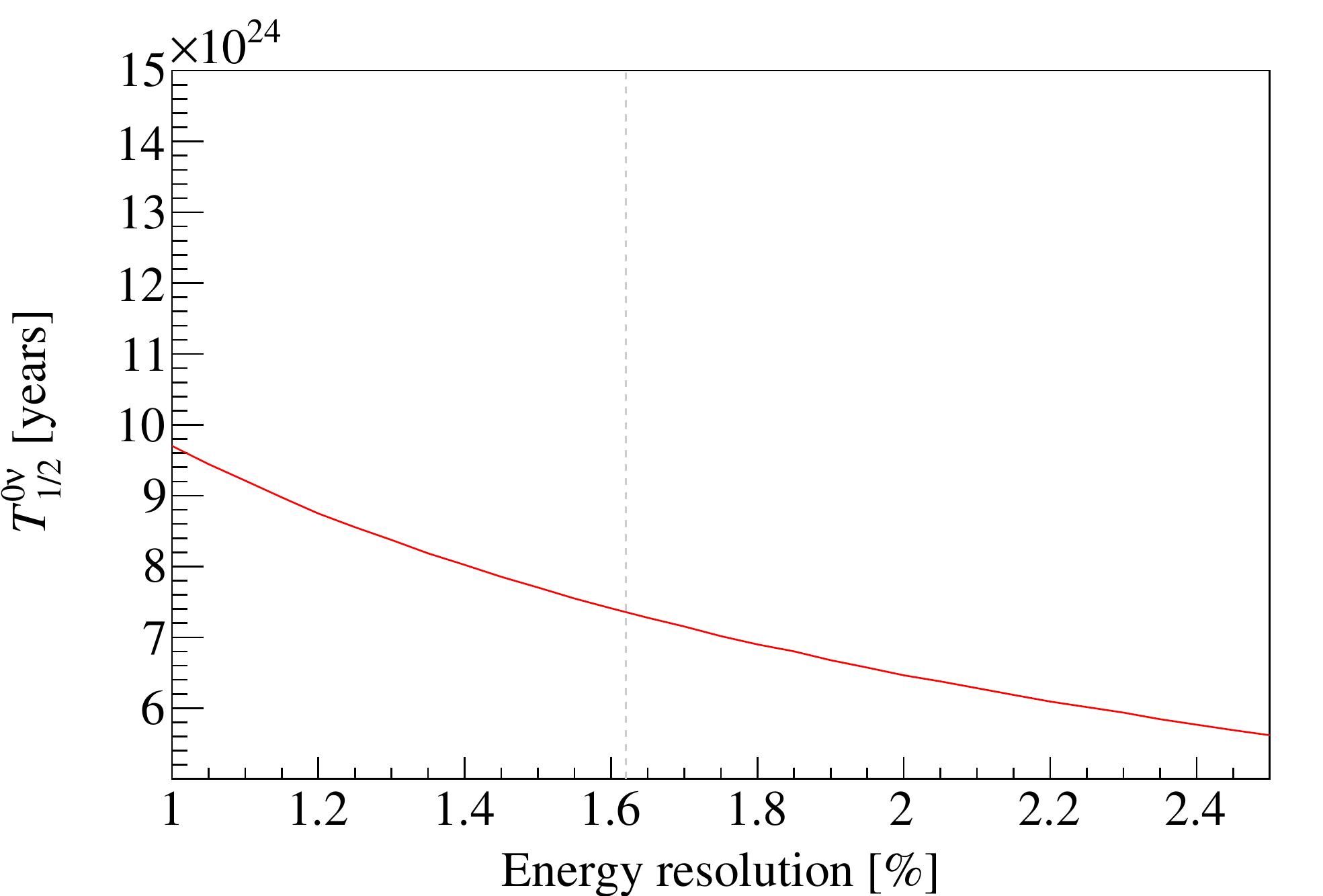}
  \caption{Dependence of the sensitivity to the \tntb\ decay of \xf\ with the level of \ke\ contamination in LXe (left), and dependence of the sensitivity to the \zntb\ decay of \xf\ with the energy resolution at $Q=$ 825.8 keV (right). The gray dashed line indicates the values assumed in this work, namely 0.3 ppt g/g $^{nat}$Kr/Xe and 1.62\%, respectively.}
  \label{kryp_resol}
\end{figure*}

As discussed in Sec. \ref{backgrounds}, the search for the \tntb\ decay of \xf\ could be severely affected by the background from \ke\ decays, given that it is one of the most important contributions in the SR and its energy spectrum is similar to that of signal. The impact of this background is assessed by calculating the sensitivity to the \tntb\ decay as a function of the \ke\ contamination in LXe. The results are shown in Fig. \ref{kryp_resol}, and indicate that such sensitivity would increase to 3$\times$10$^{24}$ years at 90\% CL if the actual \ke\ activity is twenty times smaller than the value assumed here. In this case, the LZ observation potential for \ttn\ would increase to 1.7$\times$10$^{24}$ years at three-sigma level.

The \zntb\ decay signal consists of a single line at $Q=$ 825.8 keV, and therefore the sensitivity to this process could differ from the prediction above if the actual energy resolution of the experiment departs from the 1.64\% value assumed here. The impact of this effect is assessed by calculating the dependence of the sensitivity with the energy resolution, see Fig. \ref{kryp_resol}. It is found that the decrease in sensitivity will be small if the actual energy resolution is slightly worse than the assumed value. For example, the sensitivity would drop to 6.9$\times$10$^{24}$ years if the energy resolution were 1.8\%.

The sensitivity obtained for the \zntb\ decay is used to determine the potential of LZ to constrain the absolute scale of the neutrino masses, based on Eq. \ref{eq_0nu2beta}. The value of $G_{0\nu}$ is set to 7.61$\times$10$^{-16}$ (years)$^{-1}$~\cite{yale, kotila2012}, assuming that the axial vector coupling constant $g_A$ is equal to 1.269. The value of $M_{0\nu}$ depends on the nuclear model considered, being 4.05 and 4.12 for IBM-2 and QRPA, respectively~\cite{barea2015, simkovic2013}. By setting \tzn\ equal to the median limit calculated above, the sensitivity to $\langle m_{\beta\beta}\rangle$ is found to be 1.04 eV and 1.02 eV for IBM-2 and QRPA, respectively. This result is about a factor five above the current best limit obtained by the KamLAND-Zen experiment~\cite{kamlandzen2016} and the limit expected for LZ~\cite{lz2020a}, both based on the \zntb\ decay of \xs.

If the \zntb\ decay of \xs\ were observed then the measurement of \tzn\ of \xf\ would allow to obtain $R_{0\nu}(\mathrm{^{136}Xe}, \mathrm{^{134}Xe})$. Using Eq. \ref{eq_R_0nu}, and given the existing constraints on \tzn\ of \xs, it is found that only values of \tzn\ of \xf\ above 2.3$\times$10$^{26}$ years are compatible with the values of $R_{0\nu}(\mathrm{^{136}Xe}, \mathrm{^{134}Xe})$ quoted in Sec. \ref{introduction}. This calculation assumes the values of $G_{0\nu}$ provided by the Nuclear Theory group at Yale University~\cite{yale, kotila2012}. The sensitivity of LZ to the \zntb\ decay of \xf\ is below this limit on \tzn, and therefore it would not be possible to determine $R_{0\nu}(\mathrm{^{136}Xe}, \mathrm{^{134}Xe})$ with this experiment.

\begin{figure*}
  \includegraphics[width=0.48\textwidth]{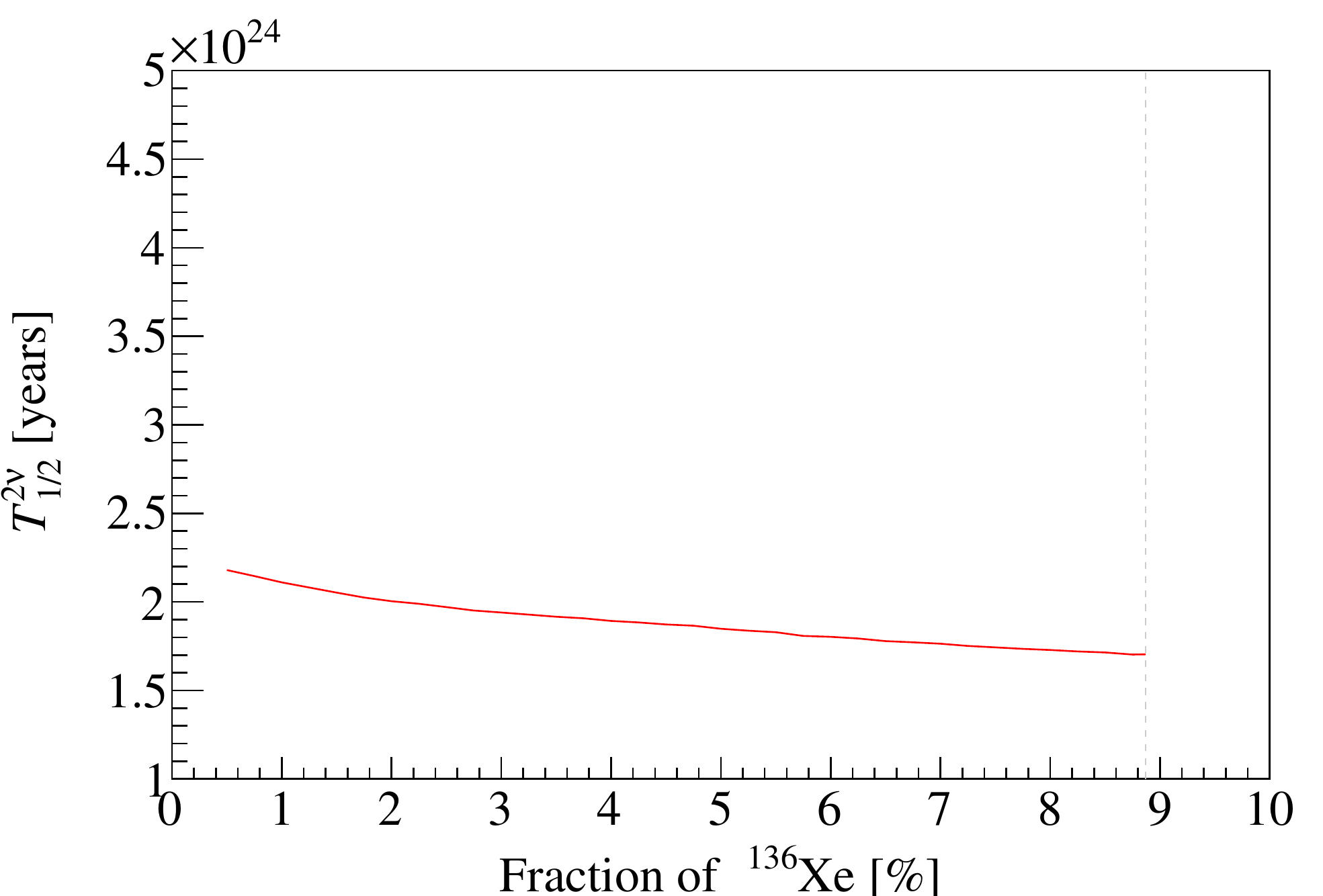}
  \includegraphics[width=0.48\textwidth]{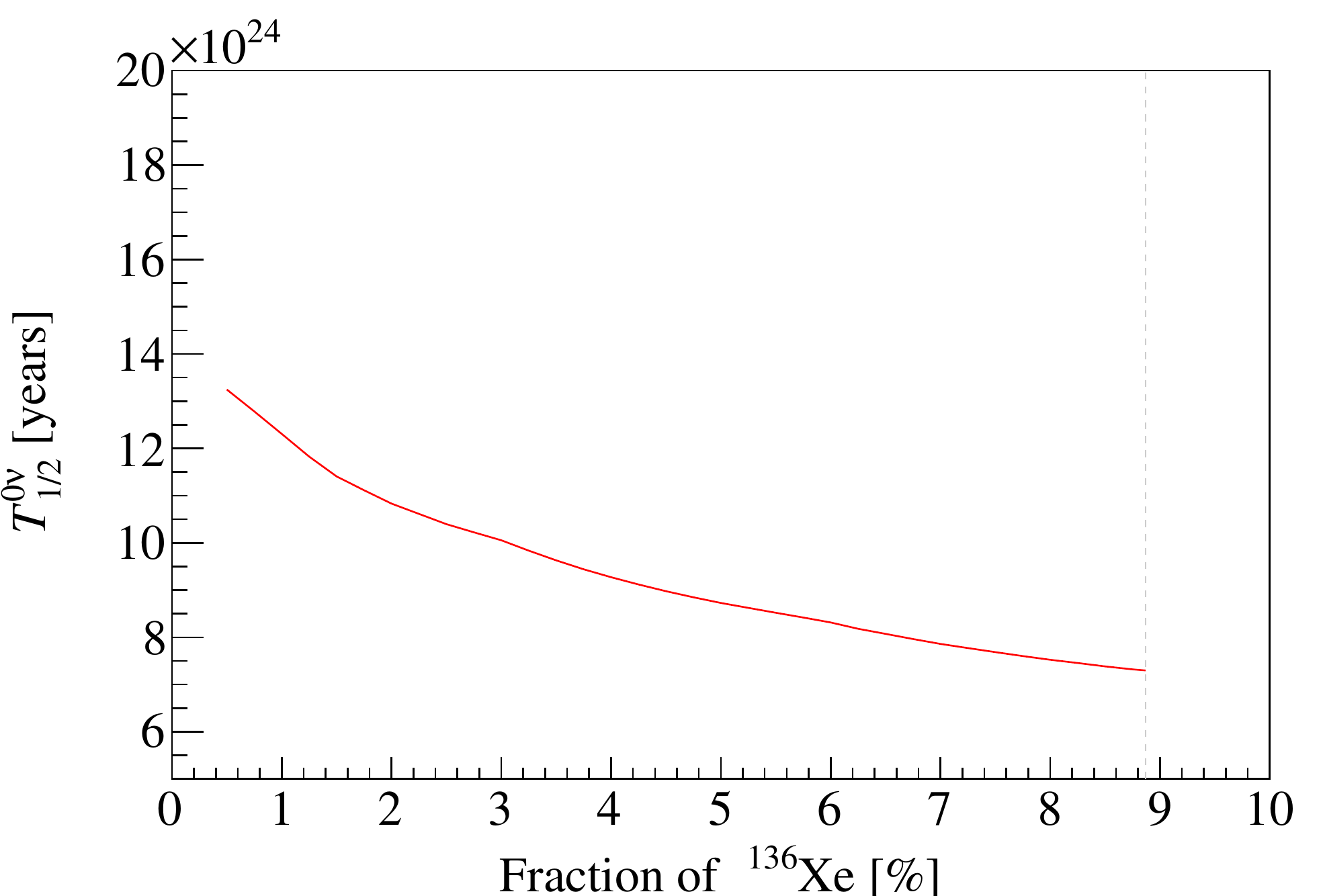}
  \caption{Dependence of the sensitivity to the \tntb\ (left) and \zntb\ (right) decays of \xf\ with the isotopic abundance of \xs. The gray dashed line indicates the natural isotopic abundance of \xs, namely 8.87\%.}
  \label{xe136}
\end{figure*}

If an opportunity arises to enrich xenon in \xs\ to search for \zntb\ with this isotope, the remaining part of xenon would be depleted in \xs, and could be used to study the decay of \xf\ with reduced background levels. This depletion would also favor the DM searches in LZ as \xs\ is an important background for these analyses. The dependence of the \xf\ decay sensitivity on the isotopic abundance of \xs\ is shown in Fig. \ref{xe136}, assuming that the relative abundances among the other isotopes remain unchanged. In particular, if the isotopic abundance of \xs\ could be lowered to 1\%, the sensitivity to the \tntb\ and \zntb\ decays would improve to 2.1$\times$10$^{24}$ and 1.2$\times$10$^{25}$ years, respectively.

\section{Conclusions}\label{conclusions}
The sensitivity of the LZ experiment to the \tntb\ and \zntb\ decays of \xf\ has been presented, assuming a live time of 1000 days. This experiment, primarily designed to search for DM particles, consists of a large detector of natural xenon with very low background levels, and therefore provides an exceptional opportunity to also search for these rare decays. The sensitivities have been calculated using the PLR formalism, considering only the information of the energy spectrum in an optimal FV.

LZ has the potential to exclude values of \ttn\ up to 1.7$\times$10$^{24}$ years at 90\% CL, and observe values of \ttn\ up to 8.7$\times$10$^{23}$ years at three-sigma level, therefore surpassing the current best limit~\cite{exo2017} by more than three orders of magnitude, and reaching the domain of the predictions provided by nuclear models. If the \ke\ contamination in LXe is reduced by a factor of twenty with respect to the current LZ requirement of 0.3 ppt g/g $^{nat}$Kr/Xe, it would be possible to observe values of \ttn\ up to 1.7$\times$10$^{24}$ years at three-sigma level.

LZ has the potential to exclude values of \tzn\ up to 7.3$\times$10$^{24}$ years at 90\% CL, improving the current best limit~\cite{exo2017} by almost two orders of magnitude.

\section*{Acknowledgements}
The research supporting this work took place in whole or in part at the Sanford Underground Research Facility (SURF) in Lead, South Dakota. Funding for this work is supported by the U.S. Department of Energy, Office of Science, Office of High Energy Physics under Contract Numbers DE-AC02-05CH11231, DE-SC0020216, DE-SC0012704, DE-SC0010010, DE-AC02-07CH11359, DE-SC0012161, DE-SC0014223, DE-SC0010813, DE-SC0009999, DE-NA0003180, DE-SC0011702, DESC0010072, DE-SC0015708, DE-SC0006605, DE-SC0008475, DE-FG02-10ER46709, UW PRJ82AJ, DE-SC0013542, DE-AC02-76SF00515, DE-SC0018982, DE-SC0019066, DE-SC0015535, DE-SC0019193, DE-AC52-07NA27344, and DOE-SC0012447. This research was also supported by U.S. National Science Foundation (NSF); the U.K. Science \& Technology Facilities Council under award numbers ST/M003655/1, ST/M003981/1, ST/M003744/1, ST/M003639/1, ST/M003604/1, ST/R003181/1, ST/M003469/1, ST/S000739/1, ST/S000666/1, ST/S000828/1, and ST/S000879/1 (JD); Portuguese Foundation for Science and Technology (FCT) under award numbers PTDC/FIS-PAR/28567/2017; the Institute for Basic Science, Korea (budget numbers IBS-R016-D1). We acknowledge additional support from the STFC Boulby Underground Laboratory in the U.K., the GridPP~\cite{gridpp2005, gridpp2009} and IRIS Consortium, in particular at Imperial College London and additional support by the University College London (UCL) Cosmoparticle Initiative. This research used resources of the National Energy Research Scientific Computing Center, a DOE Office of Science User Facility supported by the Office of Science of the U.S. Department of Energy under Contract No. DE-AC02-05CH11231. The University of Edinburgh is a charitable body, registered in Scotland, with the registration number SC005336. The assistance of SURF and its personnel in providing physical access and general logistical and technical support is acknowledged.

\bibliography{LZ_sensitivity_2beta_Xe134}
\end{document}